\documentclass[12pt]{article}
\usepackage{axodraw}
\usepackage{colordvi}
\usepackage{latexsym}
\usepackage{epsfig}
\begin{document}
\begin{center} {\Large \bf  Towards a Consistent SUSY QFT\\[0.5cm]
 in Extra Dimensions
} \vspace{1cm}

{\large \bf D.I.Kazakov} \vspace{0.7cm}

{\it Bogoliubov Laboratory of Theoretical Physics, Joint
Institute for Nuclear Research, Dubna, Russia \\[0.2cm] and\\[0.2cm]
Institute for Theoretical and Experimental Physics, Moscow,
Russia}
\end{center}

\begin{abstract}
We consider N=1 SUSY gauge theory in six dimensions in components
and show that provided the Dynkin indices of the matter fields
representations satisfy the relation $\sum T(R)= C_2(G)$, the
gauge sector is completely one-loop finite. In the  matter sector
the UV divergences form several invariant structures some of which
are cancelled in physical amplitudes.  Thus, the theory which  is
in general non-renormalizable may be consistent and even finite.
Consequences for the SUSY GUT models in the bulk are briefly
discussed.
\end{abstract}

\section{Introduction}
It became  nowadays popular to consider theories in extra
dimensions as possible candidates for models of physics beyond the
Standard Model. They provide new scenarios for the coupling
unification as well as are able to elegantly solve some problems
like doublet-triplet splitting, suppression of proton decay, SUSY
breaking, etc. (See e.g. Ref.\cite{Alt} and references therein.)
Usually, for the sake of simplicity one considers  one extra
dimension and then assumes compactification on the orbifold.
Particular  models with $S^1/(Z_2\times Z_2')$ compactification
are shown to possess some interesting properties and may serve as
a basis for the Grand Unified Theories ~\cite{Alt,BHN}. In this
case, one has N=1 supersymmetry in a five-dimensional bulk which
is equivalent to N=2 SUSY at a four dimensional brane. The field
content of the resulting theory at the brane depends on the
compactification prescription and on quantum numbers with respect
to the orbifold symmetries adjusted to the fields. Thus, below the
compactification scale (which might be the GUT scale), one has the
resulting D=4 theory on the brane with specific properties and
above this scale one has a full N=1 D=5 theory in the bulk.

One may wonder whether this extra dimensional theory can be
considered as a consistent QFT in any sense. Since by general
power counting it is non-renormalizable, it looks hardly possible.
However, there is a chance that all the UV divergences cancel each
other, like it takes place in N=4,\ 2 and even N=1 SUSY theories
in D=4 ~\cite{finite}, and one has a consistent theory.

One way to consider an extra dimensional theory is the
Kaluza-Klein approach. In this case, one takes the Fourier
transform over the extra dimensions and obtains an infinite tower
of states with quantized masses. Then one has to sum over all the
states. This sum is usually divergent and a special prescription
is needed to regularize it. Following this approach divergences in
D=5 SUSY theory have been studied  in~\cite{Ant,GN,KK}  for  the
scalar effective potential. Some cancellations of UV divergences
have been found.

The detailed structure of the K-K modes depends on the
compactification pattern. Provided that in the zero mode sector
the divergences cancel each other, one may wonder if this is also
possible at each floor of an infinite tower. This way one may get
a finite theory.

In what follows, we investigate the other  possibility and
consider explicitly D=6 N=1 SUSY gauge theory. We  show that
indeed under certain circumstances UV divergences may cancel each
other and one can have a totally finite consistent quantum field
theory in extra dimensions. These models are very distinguished by
their properties and may serve as a basis for SUSY GUT models
mentioned above. Below we discuss some of their properties.

\section{The Model}

We consider D=6 gauge QFT  with on shell  N=1 supersymmetry
formulated in components. D=6 is chosen for simplicity as the
lowest even dimension. It has the same N=1 supersymmetry as the
D=5 one (equivalent to N=2 SUSY in D=4), but the integration in
even dimensions is more familiar (in odd dimensions there are no
one loop divergences in dimensional regularization which we are
going to apply in the calculations). We do not take any particular
compactification pattern, since the UV divergences do not depend
on it; they reflect the small distance properties where locally
one has the flat Minkowski metric. We assume that the theory is
regularized in a SUSY invariant way but for practical purposes we
take the dimensional regularization (or reduction), since there is
no difference in the one loop order as concerns the UV
divergences.

In what follows, we take the usual gauge invariant Lagrangian for
the gauge and matter fields, and choose the background field gauge
as being more useful for the calculations. Of course, the
superfield formalism~\cite{superf} would be most appropriate  for
our purposes and one should try to apply it. However, in this
paper we confine ourselves to the component approach as a more
familiar one. Then, the gauge and N=1 SUSY invariant Lagrangian in
D=6 is~\cite{Fayet}
\begin{eqnarray}
  {\cal L} & = &  -\frac 14Tr\ F_{\mu\nu}F^{\mu\nu}+ i\bar \lambda\hat D\lambda
  +i\bar \psi\hat D\psi + (D_\mu\phi_1)^+(D_\mu\phi_1)
  +(D_\mu\phi_2)^+(D_\mu\phi_2)\label{L}\\
  & + &  i\sqrt 2g[(\bar \psi \lambda\phi_1+\bar \lambda\psi\phi_1^+)+
  (\bar \psi \lambda^c\phi_2+\bar \lambda^c\psi\phi_2^+)]-
  \frac{g^2}{2}|\phi_1^\dagger T^a\phi_1-\phi_2^\dagger
  T^a\phi_2|^2+2g^2|\phi_1^\dagger T^a\phi_2|^2
  \nonumber
\end{eqnarray}
and contains the following set of fields: one gauge field,
$A_\mu^a$, one Weyl gaugino field, $\lambda$, a set of chiral
matter fields in a representation $R$, $\psi$, and the
corresponding complex scalar fields $\phi_1$ and $\phi_2$.

\subsection{The gauge sector}
In the gauge sector,  due to the background field gauge invariance
the divergent structures in the one loop order can take one of the
following forms:
\begin{eqnarray}
I_1&=&Tr D_\rho F_{\mu\nu} D_\rho F_{\mu\nu}\ , \label{str} \\
I_2 &=&Tr D_\mu F_{\mu\nu} D_\rho F_{\rho\nu}\ , \nonumber \\
I_3&=&  Tr D_\rho F_{\mu\nu} D_\mu F_{\rho\nu}\ , \nonumber \\
I_4&=& Tr F_{\mu\nu} F_{\nu\rho} F_{\rho\mu}\ . \nonumber
\end{eqnarray}
However, these invariants are not independent. Due to the relation
$[D_\mu,D_\nu]=F_{\mu\nu}$, and the Bianchy identity $D_\mu
F_{\nu\rho}+D_\rho F_{\mu\nu}+D_\nu F_{\rho\mu}=0$ one has
only 2 independent structures and  can choose any of them. We take
the first two. Then calculating the diagrams and extracting the
contribution to two independent Lorentz structures one can find
the coefficients in front of them.

The structures written above contain 2-,3-,4-,5- and 6-leg
diagrams. For simplicity, we consider 2- and 3-point functions. We
use the Feynman rules from Ref.\cite{abbot}. The calculation of
the two-point function can be performed exactly in an arbitrary
dimension. The result is proportional to the transverse tensor and
the proper power of momenta. The coefficient functions are
summarized in the table (in Feynman gauge).
\begin{table}[h]
  \centering
  \begin{tabular}{|c|c|l|c|c|}\hline &&&& \\
    The Diagram & Group & D-dim Expression& $D=$& $D=$\\
    &factor&&$4-2\varepsilon$& $4-2\varepsilon$\\
    \hline
 Gauge loop\phantom{\LARGE D}
     & $C_2(G)$ &$(-)^{[D/2]+1}\frac{\Gamma(2-D/2)\Gamma^2(D/2)}{\Gamma(D)}2
  \frac{7D-8}{D-2}$
      & $-\frac{\displaystyle 10}{\displaystyle 3\varepsilon}$ & $
     -\frac{\displaystyle 17}{\displaystyle 30\varepsilon}$\\ \hline
Ghost loop\phantom{\LARGE D}
     & $C_2(G)$ &$(-)^{[D/2]+1}\frac{\Gamma(2-D/2)\Gamma^2(D/2)}{\Gamma(D)}
  \frac{4}{D-2} $
     & $ -\frac{\displaystyle 1}{\displaystyle 3\varepsilon}$ &$
      -\frac{\displaystyle 1}{\displaystyle 30\varepsilon}$ \\
      \hline (Complex) Scalar\phantom{\LARGE D}
    & $T(R)$ & $(-)^{[D/2]}\frac{\Gamma(2-D/2)\Gamma^2(D/2)}{\Gamma(D)}
  \frac{4}{D-2}$
     &$  \frac{\displaystyle 1}{\displaystyle 3\varepsilon}$ &$
     \frac{\displaystyle 1}{\displaystyle 30\varepsilon}$ \\  \hline
    Majorana (Weyl) fermion & $T(R)$ &$ (-)^{[D/2]}
  2^{[D/2]}\frac{\Gamma(2-D/2)\Gamma^2(D/2)}{\Gamma(D)}$\phantom{\LARGE D}
       & $ \frac{\displaystyle 2}{\displaystyle 3\varepsilon}$& $
     \frac{\displaystyle 8}{\displaystyle 30\varepsilon} $\\
  \hline
  \end{tabular}
  \caption{The two-point gauge function}\label{1}
\end{table}

It is instructive to consider separately the dimension of
integration $D$ and the  space-time dimension $D'$. The latter
corresponds to Lorentz algebra. Then the formulas in the table are
modified as follows: in the first line one has to change
\begin{equation}\label{add}
7D-8 \ \Rightarrow \ 8(D-1)-D' +
\frac{(D-4)(D-1)\alpha(8-\alpha)}{ 2(D-2)},
\end{equation}
where we give the result in an arbitrary $\alpha$-gauge
($\alpha=0$ corresponds to Feynman gauge), and in the last line
one has to change $2^{[D/2]}$ by $2^{[D'/2]}$.

Let us first take the $D=D'=4, \ N=1$ case. It corresponds to a
gauge and a ghost field, one Majorana spinor in adjoint
representation, and a number of supermultiplets  which contain a
Weyl spinor and a complex scalar in representation R. Then, the
pole term is
$$ -\frac{11}{3}C_2(G)+\frac{2}{3}C_2(G)+\frac 23 T(R)+\frac 13 T(R)=
-3C_2(G)+T(R),$$ i.e., cancellation of divergences is possible if
\begin{equation}
\sum T(R) = 3 C_2(G). \label{div}
\end{equation}
In the $D=D'=4, \ N=2$ case, one has to take  one supermultiplet
in the adjoint representation and add a mirror partner to any
matter supermultiplet. The  pole term here is
$$ -\frac{11}{3}C_2(G)+\frac{2}{3}C_2(G)+\frac{2}{3}C_2(G)+ \frac{1}{3}C_2(G)+2(
 \frac 23 T(R)+\frac 13 T(R))= -2C_2(G)+2T(R).$$
Hence, on has instead of (\ref{div})
\begin{equation}
\sum T(R) =  C_2(G). \label{div2}
\end{equation}
This case corresponds to the zero modes of D=6 theory in the K-K
approach. So for zero modes the UV divergences in the gauge sector
cancel if eq.(\ref{div2}) is satisfied. If the other modes follow
the same pattern, one can get a finite theory.

 Consider now the $D=D'=6, \ N=1$ case. This corresponds to one
Weyl spinor in the adjoint representation and matter
supermultiplets in R representation which contain a Weyl spinor
and two complex scalars. The pole term, which corresponds to
logarithmic divergence, is (in Feynman gauge)
$$ -\frac{17}{30}C_2(G)-\frac{1}{30}C_2(G)+\frac{8}{30}C_2(G)+ \frac{8}{30}T(R)
+\frac{2}{30} T(R)= -\frac 13 C_2(G)+ \frac 13 T(R),$$ i.e. one
has the same relation (\ref{div2}) as in the D=4 case.

However, there is also a quadratic divergence in D=6. The
corresponding gauge invariant operator in this case is simply $Tr\
F_{\mu\nu}F_{\mu\nu}$. It can be reproduced in dimensional
regularization as the residue at the pole at $\varepsilon =1$, i.e
one has to take $D=4$ . Substituting in eq.(\ref{add}) $D'=6, D=4$
one gets for the quadratic divergence
$$\frac{1}{6}[(-18-2+8)C_2(G)+12T(R)]
=2[-C_2(G)+T(R)],$$ i.e. one has again eq.(\ref{div2}), but the
gauge dependence disappears here from the result. It is gauge
invariant.

Though we are interested here in D=6 theory, it is interesting to
apply eq.(\ref{add}) to D=10 case, since it corresponds to N=4
SUSY theory in D=4.  One has here besides logarithmic also
quadratic, quartic and sextic divergences. They can be reproduced
from eq.(\ref{add}) by allowing $D'=10$ and $D=10,8,6$ and $4$,
respectively. In this case one has no matter, but one
Majorana-Weyl spinor in adjoint representation. The pole term is
proportional to (for any D)
$$C_2(G) [-16+16 +\frac{(D-4)(D-1)\alpha(8-\alpha)}{2(D-2)}],$$
i.e. all divergences cancel in Feynman gauge and the highest
divergence is gauge invariant and vanishes in any gauge.

Consider now the three-point vertices. The divergent part of  the
3-point diagrams in the $D=6,\ N=1$ case is (in Feynman gauge):
\begin{eqnarray*}
\frac{T(R)-C_2(G)}{6} \Big\{&{\displaystyle \frac 43}&\left.
k_{1\mu} k_{1\nu} k_{1\rho} - \frac 43 k_{2\mu} k_{2\nu}
k_{2\rho}+ \frac 23 k_{1\rho} k_{1\mu} k_{2\nu} - \frac 23
k_{1\mu} k_{2\rho} k_{2\nu}\right. \\ &+& \left. \frac 43 k_{1\mu}
k_{1\nu} k_{2\rho}- \frac 43 k_{2\mu}
k_{2\nu} k_{1\rho} \right.\\
&+& g_{\mu\nu} k_{1\rho}\left[\frac 23 k_1^2 + 2 k_2^2 + \frac 43
(k_1k_2)\right]- g_{\mu\nu} k_{2\rho}\left[\frac 23 k_2^2 + 2
k_1^2
 +\frac 43 (k_1 k_2)\right]\\
&- &g_{\mu\rho} k_{1\nu}\left[\frac 83 k_1^2 + \frac 83 k_2^2
+\frac 83
(k_1 k_2)\right] - g_{\mu\rho} k_{2\nu}\left[\frac 23 k_2^2 + \frac 43 k_1^2\right] \\
&+& \left.g_{\nu\rho} k_{1\mu}\left[\frac 23 k_1^2 + \frac 43
k_2^2\right] + g_{\nu\rho} k_{2\mu}\left[\frac 83 k_1^2 + \frac 83
k_2^2 + \frac 83 (k_1 k_2)\right] \right\}\ .
\end{eqnarray*}
This leads to the following terms (after reduction)
\begin{equation}
\frac{T(R)-C_2(G)}{3} f^{abc} \left(2\ \partial_\mu \partial_\rho
A_\mu^a
\partial_\nu A_\nu^b A_\rho^c
 +2\ \partial^2 \partial_\nu A_\mu^a A_\mu^b
A_\nu^c + 6\ \partial_\nu A_\mu^a \partial^2 A_\mu^b A_\nu^c + 4\
\partial_\nu \partial_\rho A_\mu^a \partial_\rho A_\mu^b A_\nu^c
\right)\label{res}
\end{equation}
At the same time, expansion of the invariants over the fields up
to the third oder gives
\begin{eqnarray*} I_1 & = &
2A_\mu^a\partial^2(g_{\mu\nu}\partial^2-\partial_\mu\partial_\nu)
A_\nu^a +f^{abc}(4\partial^2\partial_\nu A_\mu^a A_\mu^b A_\nu^c
-4\partial_\rho\partial_\nu A_\mu^a \partial_\nu A_\mu^b A_\rho^c
+4\partial_\rho\partial_\nu A_\mu^a \partial_\mu A_\nu^b A_\rho^c),\\
I_2 & = &
A_\mu^a\partial^2(g_{\mu\nu}\partial^2-\partial_\mu\partial_\nu)
A_\nu^a +f^{abc}(2\partial^2\partial_\nu A_\mu^a A_\mu^b A_\nu^c -
6\partial^2A_\mu^a\partial_\nu A_\mu^b A_\nu^c
+4\partial_\rho\partial_\nu A_\mu^a\partial_\nu A_\mu^b A_\rho^c\\
&&+2\partial_\mu\partial_\nu A_\mu^a \partial_\rho A_\nu^b
A_\rho^c).
\end{eqnarray*}
Adding them together one finds
\begin{eqnarray*}
&&xI_1+yI_2
=A_\mu^a\partial^2(g_{\mu\nu}\partial^2-\partial_\mu\partial_\nu)
A_\nu^a (2x+y)\\
&&+2f^{abc}\partial^2\partial_\nu A_\mu^a A_\mu^b A_\nu^c (2x+y)
-2f^{abc}\partial_\rho\partial_\nu A_\mu^a \partial_\nu A_\mu^b
A_\rho^c
(2x-2y)\\
&&-2f^{abc}\partial^2A_\mu^a\partial_\nu A_\mu^b A_\nu^c
(3y)+2f^{abc}\partial_\mu\partial_\rho A_\mu^a \partial_\nu
A_\nu^b A_\rho^c (2x+y)-2f^{abc}\partial_\rho A_\mu^a \partial_\mu
A_\nu^b\partial_\nu A_\rho^c(2x).
\end{eqnarray*}
Comparing this with eq.(\ref{res}), one gets ($C_2(G)\equiv C_A$)
$$2x+y=\frac{T_R-C_A}{3}, \ \ 2x-2y=-2\frac{T_R-C_A}{3}, \ \
 3y=T_R-C_A, \ \ 2x=0.$$
Thus, one has the one-loop logarithmic divergences in the form
\begin{equation}
\frac{T_R-C_A}{3}\ Tr D_\mu F_{\mu\nu} D_\rho
F_{\rho\nu}.\label{gauge}
\end{equation}
One finds that the result for ALL the structures is proportional
to $\sum T(R) - C_2(G)$, like  for the two-point functions, and
vanishes if eq.(\ref{div2}) is satisfied. Due to the fact that all
the structures vanish we claim that all the one loop divergences
in the gauge sector cancel for $\sum T(R) = C_2(G)$!

The situation is not that simple in an arbitrary $\alpha$-gauge.
Equation (\ref{gauge}) in accordance with (\ref{add}) in this case
looks like
\begin{equation}
\frac{T_R-C_A(1+\alpha-\alpha^2/8)}{3}\ Tr D_\mu F_{\mu\nu} D_\rho
F_{\rho\nu}.\label{gauge2}
\end{equation}
The cancellation is not obvious anymore. One has to consider the
proper combination of the Green functions to observe the
cancellation of the gauge dependence and associated cancellation
of the UV divergences. We come back to this problem when
considering the matter sector.

\subsection{The matter sector}

 In the matter sector, in four
dimensions one has both the propagators and the vertices to
diverge and only the proper combination of them is finite. The
situation is even more complicated in D=6, since here one has
extra powers of momenta in the diagrams and the usual cancellation
does not work. Still one can try to find a proper combination of
vertices that gives finite matrix elements. We consider some
examples of this cancellation below but first we calculate the
one-loop diagrams with the matter fields.
 \vspace{0.3cm}

 \underline{The spinor fields}\vspace{0.3cm}

Consider the fermions first. Restricting oneself to the diagrams
with two fermion legs, in the one-loop order one has the following
invariants:
\begin{eqnarray}
  J_1 & = &  \bar \Psi \gamma^\nu D_\mu D_\mu D_\nu \Psi\ , \nonumber \\
  J_2 & = &  \bar \Psi \gamma^\nu D_\mu D_\nu D_\mu \Psi\ , \nonumber\\
  J_3 & = &  \bar \Psi \gamma^\nu D_\nu D_\mu D_\mu \Psi\ , \nonumber\\
  J_4 & = &  \bar \Psi \gamma^\mu \gamma^\nu \gamma^\rho D_\mu D_\nu D_\rho
  \Psi\ . \label{spinor}
\end{eqnarray}
Expanding them up to the third order over the gauge fields one has
\begin{eqnarray*}
  J_1& = &  \bar \Psi \gamma^\mu \left(\partial^2\partial_\mu +
  2A_\nu \partial_\nu\partial_\mu + \partial_\nu A_\nu \partial_\mu
  +\partial^2 (A_\mu \ \Psi\right)\\
  && \Rightarrow p^2\hat p + (2\hat p_1 p_1^\mu +
  \hat p_1 p_3^\mu + p_2^2\gamma^\mu) A_\mu, \\
  J_2& = &  \bar \Psi \gamma^\mu \left(\partial^2\partial_\mu +
  A_\nu \partial_\nu\partial_\mu + \partial_\nu A_\mu \partial_\nu
  +A_\mu \partial_\nu\partial_\nu
  +\partial_\nu\partial_\mu (A_\nu \ \Psi\right)\\
  && \Rightarrow p^2\hat p + (\hat p_1 p_1^\mu +
  \hat p_2 p_2^\mu  + p_1^2\gamma^\mu  + p_1p_3 \gamma^\mu) A_\mu, \\
 J_3& = &  \bar \Psi \gamma^\mu \left(\partial^2\partial_\mu +
  A_\mu \partial^2 + 2\partial_\mu A_\nu \partial_\nu
  +2A_\nu \partial_\nu\partial_\mu
  +\partial_\nu\partial_\mu A_\nu +\partial_\nu A_\nu \partial_\mu
  \right)\Psi\\ &&
  \Rightarrow p^2\hat p + (2\hat p_3 p_1^\mu  +
 2\hat p_1 p_1^\mu  + \hat p_3 p_3^\mu + \hat p_1 p_3^\mu
 + p_1^2\gamma^\mu) A_\mu,\\
 J_4&=&\bar \Psi \gamma^\mu \left(\partial^2\partial_\mu +
   \partial^2(A_\mu + \hat \partial( A_\mu \hat \partial
  +A_\mu \hat \partial \hat \partial
  \right)\Psi\\ &&
\Rightarrow p^2\hat p + (p_2^2\gamma^\mu  -
 \hat p_2 \gamma^\mu  \hat p_1
  + p_1^2\gamma^\mu) A_\mu.
\end{eqnarray*}
Adding these expressions together one finds ($p_1+p_2+p_3=0$)
\begin{eqnarray*}
&&  xJ_1+yJ_2+zJ_3+tJ_4= \hat p p^2 (x+y+z+t)  + A_\mu \left[ \hat
p_1 p_1^\mu (x+y) \right.\\&+& \left. \hat p_2 p_2^\mu (y+z) +
\hat p_1 p_2^\mu (-x) + \hat p_2 p_1^\mu (-z) + \gamma^\mu
[(z+t)p_1^2-yp_1p_2+(x+t)p_2^2)]+\hat p_2\gamma^\mu\hat
p_1(-t)\right].
\end{eqnarray*}

At the same time, the calculation of two- and three-point diagrams
gives (we take here an arbitrary $\alpha$-gauge)
\begin{eqnarray*}
&-&\hat p p^2 \alpha\frac{C_R}{3}  +
\frac{C_A}{6}\Bigg[(-6+\alpha)\hat p_2\gamma^\mu \hat p_1+(2+\frac
52 \alpha-\frac 14\alpha^2)(\hat p_1 p_1^\mu+\hat p_2 p_2^\mu)
+(8+\frac 52 \alpha-\frac 14\alpha^2)\hat p_2 p_1^\mu
\\&+&(8+\frac 52\alpha-\frac 14\alpha^2)\hat p_1 p_2^\mu  +
\gamma^\mu[ (-2-\frac 72\alpha +\frac
14\alpha^2)(p_1^2+p_2^2)+(-10-5\alpha+\frac 12
\alpha^2)p_1p_2]\Bigg]A_\mu \\
&+& \frac{C_R-C_A/2}{3}\Bigg[\alpha\hat p_2\gamma^\mu \hat p_1-
2(\hat p_1 p_1^\mu+\hat p_2 p_2^\mu) -2(\hat p_2 p_1^\mu +\hat p_1
p_2^\mu)+(2-\alpha)\gamma^\mu(p_1^2+p_2^2)+4\gamma^\mu p_1p_2
 \Bigg]A_\mu .
\end{eqnarray*}
Comparing these expressions one gets
$$x=z=\frac 23C_R-\frac{C_A}{3}(5+\frac 54\alpha -\frac 18\alpha^2), \ \
y=-\frac 43C_R+\frac{C_A}{3}(7+\frac 52\alpha -\frac 14\alpha^2),
\ \ t=\frac{C_A}{3}(3) - \frac{C_R}{3}\alpha .
$$ Thus, the one loop divergences for the matter spinor fields have the
following form:
\begin{eqnarray}
  &\bar \Psi &\left\{ \ \gamma^\nu[D_\mu D_\mu D_\nu -2 D_\mu D_\nu D_\mu + D_\nu D_\mu
  D_\mu][\frac 23 C_R-\frac{C_A}{3}(5+\frac 54 \alpha
  -\frac{\alpha^2}{8})] \right.\label{mat}  \\
  & & \left.-[\gamma^\nu D_\mu D_\nu D_\mu -
  \gamma^\mu\gamma^\nu\gamma^\rho D_\mu D_\nu
  D_\rho]\frac{C_A}{3}(3) -\alpha\frac{C_R}{3}
  \gamma^\mu\gamma^\nu\gamma^\rho D_\mu D_\nu D_\rho \ \right\} \Psi.\nonumber
\end{eqnarray}
For the gaugino field they are slightly different
\begin{eqnarray}
  &\bar \lambda&\left\{ \  \gamma^\nu[D_\mu D_\mu D_\nu -2 D_\mu D_\nu D_\mu + D_\nu D_\mu
  D_\mu][\frac{C_A}{3}(-\frac 54 \alpha
  +\frac{\alpha^2}{8})-T_R]  \right. \label{mat2}\\
  & & \left. -[\gamma^\nu D_\mu D_\nu D_\mu -
  \gamma^\mu\gamma^\nu\gamma^\rho D_\mu D_\nu
  D_\rho]\frac{C_A+2T_R}{3}+\frac{C_A(1-\alpha)-T_R}{3}
  \gamma^\mu\gamma^\nu\gamma^\rho D_\mu D_\nu D_\rho \right\}\lambda \ .\nonumber
\end{eqnarray}
In the case when $\sum T(R)=C_A$ one has
\begin{eqnarray}
  &\bar \lambda&\left\{\ \gamma^\nu[D_\mu D_\mu D_\nu -2 D_\mu D_\nu D_\mu + D_\nu D_\mu
  D_\mu]\frac{C_A}{3}(-3-\frac 54 \alpha
  +\frac{\alpha^2}{8}) \right.  \\
  & & \left.-[\gamma^\nu D_\mu D_\nu D_\mu -
  \gamma^\mu\gamma^\nu\gamma^\rho D_\mu D_\nu
  D_\rho]\frac{C_A}{3}(3)-\frac{\alpha}{3}C_A
  \gamma^\mu\gamma^\nu\gamma^\rho D_\mu D_\nu D_\rho\right\}\lambda ,\nonumber
\end{eqnarray}
and for $\alpha=0$
\begin{equation}
  -C_A \bar \lambda \left\{\ \gamma^\nu[D_\mu D_\mu D_\nu - D_\mu D_\nu D_\mu + D_\nu D_\mu
  D_\mu] -
  \gamma^\mu\gamma^\nu\gamma^\rho D_\mu D_\nu
  D_\rho\right\}\ \lambda \ .
\end{equation}

No wonder, these expressions look complicated. The same is in D=4.
However there, if one takes the product of the 3-point function
and the square roots of the propagators for each leg, the
resulting invariant charge is finite and does not depend on the
gauge. One can try the same combination in D=6, but apparently it
does not work due to the complicated momentum structure of the
3-point function.

Since our final goal is to get finite and gauge invariant
observables, we consider the amplitude of a physical process. As
an example we take the Compton scattering. In this case one has
the combination of 2-,3- and 4-point functions shown in Fig.1.
\begin{figure}[ht]\vspace{0.5cm}
\begin{center}
\leavevmode
  \epsfxsize=12cm
 \epsffile{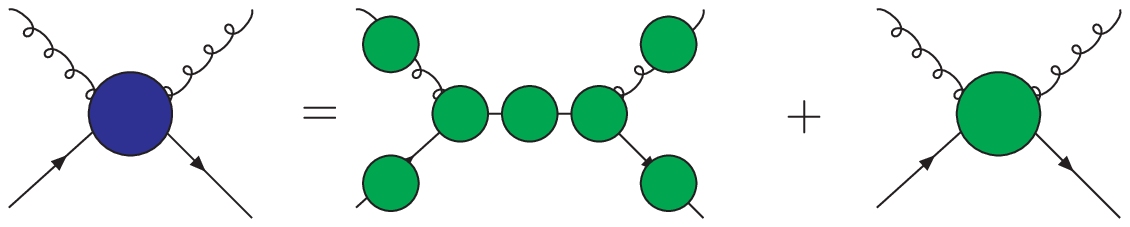}
\caption{The Compton scattering amplitude. The bulbs in the r.h.s.
denote the one-particle irreducible graphs.}
\end{center}
\end{figure}
Taking the contribution to the diagrams on the r.h.s. of Fig.1
from eq.(\ref{mat}) one gets the resulting amplitude. Since there
are three independent structures in eq.(\ref{mat}), we consider
their contribution separately. The simplest is the last one which
comes with the coefficient proportional to the gauge parameter and
is expected to disappear from the final answer. Indeed, one has
the following divergent contribution coming from this structure
\vspace{0.2cm}

\begin{center}
\leavevmode
  \epsfxsize=12cm
 \epsffile{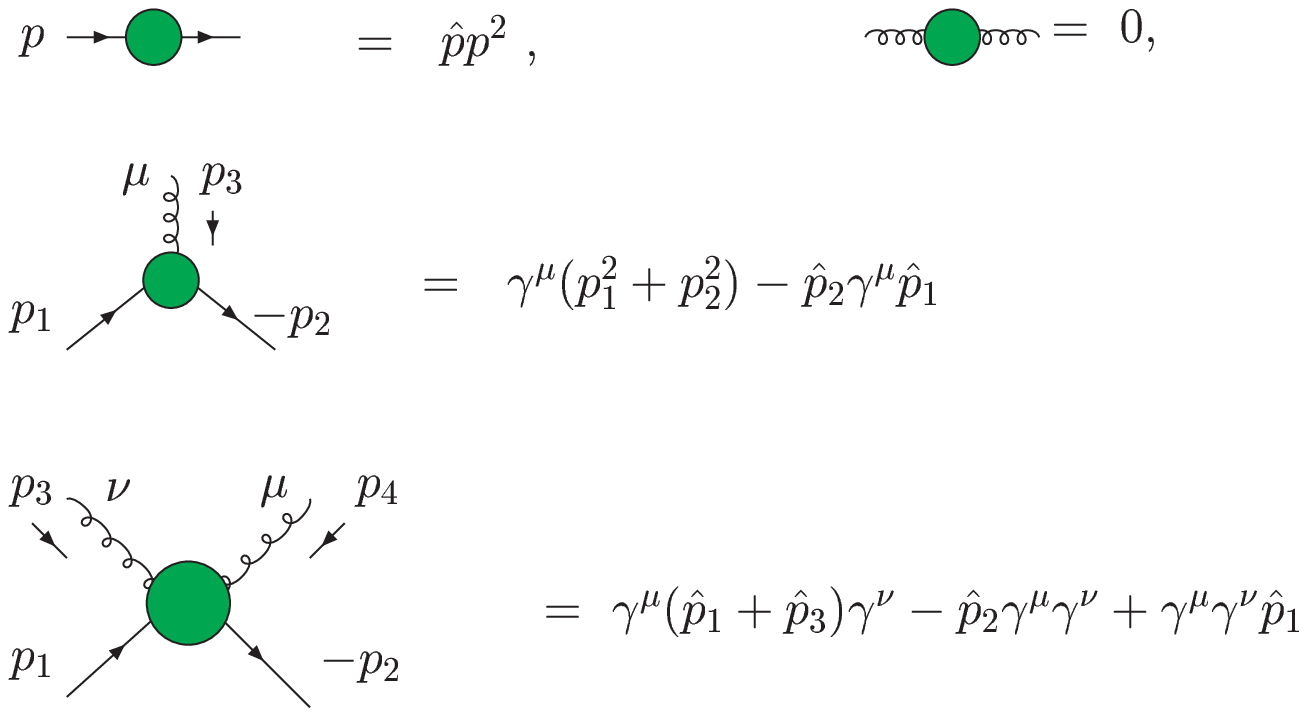}
\end{center}
which gives
\begin{eqnarray*}Amp\ =
&&i\frac{\gamma^\mu(\hat p_1+\hat
p_3)\gamma^\nu}{(p_1+p_3)^2}[p_1^2+(p_1+p_3)^2+p_2^2]+i[\gamma^\mu(\hat
p_1+\hat p_3)\gamma^\nu+ \hat
p_2\gamma^\mu\gamma^\nu-\gamma^\mu\gamma^\nu\hat p_1]\\
&-&i\frac{\gamma^\mu(\hat p_1+\hat p_3)}{(p_1+p_3)^2}[\gamma^\nu
p_1^2+\gamma^\nu(p_1+p_3)^2+(\hat p_1+\hat p_3)\gamma^\nu\hat
p_1]\\ &-&i[\gamma^\mu p_2^2+\gamma^\mu(p_1+p_3)^2-\hat
p_2\gamma^\mu(\hat
p_1+\hat p_3)]\frac{(\hat p_1+\hat p_3)\gamma^\nu}{(p_1+p_3)^2}\\
&=&0,
\end{eqnarray*}
so this contribution indeed cancels.

Unfortunately, the same is not true for the other structures
though the gauge dependence should disappear in this case as well.
We have not found yet the way how it actually happens. Strictly
speaking this should be true for the $S$-matrix which means that
one has to impose the equations of motion on the fields. However,
since we would like to trace the cancellation of the UV
divergences in all loops, we keep all the structures off shell.

\vspace{0.3cm} \underline{The scalar fields}\vspace{0.3cm}

 The situation with the scalar fields is similar. Restricting oneself again to the
diagrams with two scalar legs one has the following invariants in
the one-loop order:
\begin{eqnarray}
  S_1 & = &  (D_\mu D_\mu \Phi)^\dagger (D_\nu D_\nu \Phi),\label{sc}\\
  S_2 & = &  (D_\mu D_\nu \Phi)^\dagger (D_\mu D_\nu \Phi),\nonumber\\
  S_3 & = &  (D_\mu D_\nu \Phi)^\dagger (D_\nu D_\mu
  \Phi).\nonumber
\end{eqnarray}
Expanding them up to the third order over the gauge fields one
gets
\begin{eqnarray*}
  S_1& = & \partial^2\Phi^\dagger \partial^2\Phi+ \partial^2\Phi^\dagger\partial_\nu A_\nu\Phi
  + 2\partial^2\Phi^\dagger A_\nu\partial_\nu\Phi-
  \Phi^\dagger\partial_\mu A_\mu\partial^2\Phi -2 \partial_\mu\Phi^\dagger A_\mu
   \partial^2\Phi\\
  & \Rightarrow& p^4 +
  (p_2^2p_3^\mu+2p_2^2p_1^\mu-p_1^2p_3^\mu-2p_1^2p_2^\mu)A_\mu=
  p^4+[p_1^\mu(p_1^2+p_2^2)-p_2^\mu(p_1^2+p_2^2)]A_\mu,\\
  S_2& = & \partial_\mu\partial_\nu\Phi^\dagger \partial_\mu\partial_\nu\Phi
  + \partial_\mu\partial_\nu\Phi^\dagger\partial_\mu A_\nu\Phi
  + 2\partial_\mu\partial_\nu\Phi^\dagger A_\mu\partial_\nu\Phi-
  \Phi^\dagger\partial_\mu A_\nu\partial_\mu\partial_\nu\Phi
   -2 \partial_\nu\Phi^\dagger A_\mu\partial_\mu\partial_\nu\Phi\\
  & \Rightarrow& p^4 +
  [2p_1p_2p_2^\mu+p_2p_3p_2^\mu-2p_1p_2p_1^\mu-p_1p_3p_1^\mu]
  A_\mu =
  p^4+[p_1^\mu(p_1^2-p_1p_2)-p_2^\mu(p_2^2-p_1p_2)]A_\mu ,\\
  S_3& = & \partial_\mu\partial_\nu\Phi^\dagger \partial_\nu\partial_\mu\Phi
  + \partial_\mu\partial_\nu\Phi^\dagger\partial_\nu A_\mu\Phi
  + 2\partial_\mu\partial_\nu\Phi^\dagger A_\mu\partial_\nu\Phi-
  \Phi^\dagger\partial_\mu A_\nu\partial_\mu\partial_\nu\Phi
   -2 \partial_\nu\Phi^\dagger A_\mu\partial_\mu\partial_\nu\Phi\\
   & \Rightarrow& p^4 +
  [2p_1p_2p_2^\mu+p_2p_3p_2^\mu-2p_1p_2p_1^\mu-p_1p_3p_1^\mu]
  A_\mu =
  p^4+[p_1^\mu(p_1^2-p_1p_2)-p_2^\mu(p_2^2-p_1p_2)]A_\mu,
\end{eqnarray*}
so that the combination looks like
\begin{eqnarray*}
&&  xS_1+yS_2+zS_3= p^4 (x+y+z)\\
&& + A_\mu
[p_1^\mu(p_1^2(x+y+z)-p_1p_2(y+z)+p_2^2(x))-p_2^\mu(...)].
\end{eqnarray*}
At the same time, the calculation of the 2- and 3-point functions
gives
\begin{eqnarray*}
 &-&p^4\frac{\alpha}{2}C_R\\  &+& A_\mu \left\{p_1^\mu\left[
 \frac{C_A}{2}(p_1^2\frac{3-4\alpha}{12}+p_1p_2\frac{5-12\alpha}{6}
 -p_2^2\frac{5+15\alpha}{6})\right.\right.\\
 &+&\left.\left.(C_R-\frac{C_A}{2})(p_1^2\frac{3-\alpha}{6}+\frac 43 p_1p_2
 +p_2^2\frac{8-3\alpha}{6})-(2C_R-\frac{C_A}{2})p_1^2\frac{3+2\alpha}{12}\right]
-p_2^\mu(...)\right\}.
\end{eqnarray*}
 Comparing the above two expressions one gets
 $$x+y+z=-\frac{\alpha}{2}C_R, \ \ y+z= (\frac{1}{4}+\alpha)C_A -\frac 43 C_R, \ \
 x=-(\frac{1}{4}+\alpha)C_A +(\frac{4}{3}-\frac{\alpha}{2})C_R.$$
Thus, the one-loop divergences in the scalar sector have the form
\begin{eqnarray}
&&-[(\frac{1}{4}+\alpha)C_A -(\frac{4}{3}-\frac{\alpha}{2})C_R]
(D_\mu D_\mu \Phi)^\dagger (D_\nu D_\nu \Phi)
+y(D_\mu D_\nu \Phi)^\dagger (D_\mu D_\nu \Phi)\nonumber\\
&&+[(\frac{1}{4}+\alpha)C_A -\frac 43 C_R-y](D_\mu D_\nu
\Phi)^\dagger (D_\nu D_\mu \Phi).\label{mat3}
\end{eqnarray}
To find the value of $y$, one has to calculate the 4-point
function.

One can see that in the scalar sector, like in the spinor one,
there is no simple cancellation of divergences. One has to
consider again the proper combination of the Green functions.
Since the usual product of the propagators and the 3-point
vertices does not work here as well, we look for the physical
amplitude.

Consider again the Compton-like amplitude but for the scalar
fields and take the first invariant $S_1$ from eq.(\ref{sc}).
Then, in full analogy with the fermion case, one has
\begin{eqnarray*}
Amp &=&
i\frac{(p_1+p_3-p_2)^\mu(2p_1+p_3)^\nu}{(p_1+p_3)^2}[p_1^2+(p_1+p_3)^2+p_2^2]
-ig^{\mu\nu}(p_1^2+p_2^2)\\
&&-i\frac{(p_1+p_3-p_2)^\mu}{(p_1+p_3)^2}[p_1^\nu(p_1^2+(p_1+p_3)^2)+(
p_1+p_3)^\nu((p_1+p_3)^2+p_1^2)]\\
&&-i[(p_1+p_3)^\mu(p_2^2+(p_1+p_3)^2)-p_2^\mu(p_2^2+(p_1+p_3)^2)]
\frac{(2p_1+p_3)^\nu}{(p_1+p_3)^2}\\
&&+i[g^{\mu\nu}(p_1^2+p_2^2)+2p_1^\mu p_1^\nu+p_1^\mu
p_3^\nu-2p_2^\mu p_1^\nu-p_2^\mu p_3^\nu+2p_3^\mu
p_1^\nu+p_3^\mu p_3^\nu]\\
&=&0,
\end{eqnarray*}
i.e., this contribution drops from the amplitude. Again, like in
the spinor case, this does not happen for the other invariants,
though the gauge invariance arguments are still valid and one has
to find out how the gauge dependence actually goes away.

\subsection{The Yukawa sector}
In the Yukawa sector in D=4 due to analyticity in superspace one
has no renormalization of a superpotential. This means that all
the vertex diagrams converge and divergences can only happen in
chiral field propagators. This might also be true in D=6 in a
superfield approach, but in components it looks more complicated.
We have not studied this sector yet.

\section{SUSY GUT Models in the Bulk}

Taking eq.(\ref{div2}) seriously, one may wonder what kind of
models in the bulk satisfy this requirement. It happens to be not
so many possibilities bearing in mind that one should have at
least three generations of the SM particles.\footnote{The result
depends of course on whether or not some particles propagate in
the bulk or confined to the brane in the brane world scenario.
Here for simplicity we assume that all the particles are in the
bulk.} Looking for the values of Dynkin indices~\cite{HPS} one
finds for instance that the SU(5) theory does not satisfy
eq.(\ref{div2}), neither any other SU(N) or SO(2k+2), $k > 2$
theory does. There are only two viable models: SO(10) and E(6)
with the following particle content:
\begin{table}[h]
  \centering
  \begin{tabular}{|c|c|c|c|}\hline &&& \\
    The Model & Gauge Group & Matter fields & Higgs fields \\
    \hline
    I & SO(10)& $4 \times \underline{16}$ & -\\
    \hline
    II & SO(10)& $3 \times \underline{16}$ & $1\times \underline{16}$\\
    \hline
    III& SO(10)& $3 \times \underline{16}$ & $2\times \underline{10}$\\
    \hline
    IV & E(6) &$ 4\times \underline{27}$& - \\ \hline
    V & E(6) & $3\times \underline{27}$& $1\times\underline{27}$\\
    \hline
  \end{tabular}
  \caption{Possible consistent N=1 D=6 models}\label{2}
\end{table}

Each SM particle in these models being projected to a
4-dimensional brane obtains the mirror partner in a conjugated
representation $\bar R$, which has to be heavy enough not to be
observed.  This is the well-known problem in N=2 SUSY models in
D=4. It can be solved in the brane world scenario by adjusting
proper quantum numbers to all the particles with respect to an
orbifold symmetry group. One can then remove some unwanted
particles from the SM brane confining them to another brane,
etc~\cite{Alt,BHN}. There may be complicated scenarios when some
particles are in the bulk while the others are confined to the
brane. We do not consider these questions here, but concentrate on
the construction of a consistent QFT in extra dimensions.

\section{Conclusion}

We have demonstrated how UV divergences cancel each other in some
cases even in non-renormalizable models. In Feynman gauge it is
straightforward in the gauge sector but is rather tricky in the
matter one.

The situation can be simplified when going on shell; however,
since the equations of motion  mix the matter fields due to the
Yukawa type interactions, one should consider all the invariants
(\ref{mat},\ref{mat2},\ref{mat3}) together. This does not seem to
be simple. On the other hand, as has already been mentioned, we
would like to stay off shell to be able to go beyond one loop.

Unfortunately, we have not completed this task. We rely here on
the superfield formalism which should simplify the situation
drastically. On the other hand, the K-K approach can also be
useful if cancellation of infinite towers level by level is really
possible.

If the N=1 D=6 SUSY theory inherits some properties of N=2 D=4 one
(see e.g.\cite{S}), where the UV divergences occur only in one
loop, one can hope that these results are valid  in higher orders
as well.

\section*{Acknowledgements}
The author would like to thank A.Sheplyakov and V.Velizhanin for
help in the calculations and H.-P.Nilles, R.Nevzorov, S.Mikhailov
and A.Bakulev for valuable discussions. I am grateful to the
Theory Group of KEK where part of this work has been done, for
hospitality. Financial support from RFBR grants \# 02-02-16889 and
\# 00-15-96691 is kindly acknowledged.

\end{document}